\documentclass[11pt, a4paper]{article}

\usepackage[margin=1in]{geometry}
\usepackage[utf8]{inputenc}
\usepackage[T1]{fontenc}
\usepackage{parskip}
\usepackage{microtype}
\usepackage{graphicx}
\usepackage[most]{tcolorbox}
\usepackage{booktabs} 
\usepackage{threeparttable}
\usepackage{tabularx} 
\usepackage{multirow}
\usepackage[table]{xcolor}
\definecolor{ODIblue}{HTML}{072589}
\definecolor{text}{HTML}{222222}
\usepackage{xurl}
\usepackage{amsmath}
\usepackage{amssymb}
\usepackage{authblk}

\usepackage[colorlinks=true, linkcolor=ODIblue, citecolor=ODIblue, urlcolor=ODIblue]{hyperref}
\usepackage{natbib}

\renewcommand{\thefootnote}{\alph{footnote}}
\usepackage{tikz}

\newcommand\blfootnote[1]{%
  \begingroup
  \renewcommand\thefootnote{}\footnote{#1}%
  \addtocounter{footnote}{-1}%
  \endgroup
}

\makeatletter
\newcommand*{\samethanks}[1][\value{footnote}]{\footnotemark[#1]}
\makeatother

\title{\vspace{-1.5cm}The CitizenQuery Benchmark: A Novel Dataset and Evaluation Pipeline for Measuring LLM Performance in Citizen Query Tasks}

\author[1]{Neil Majithia\thanks{Equal Contribution. \texttt{neil.majithia@theodi.org}}}
\author[2]{Rajat Shinde\thanks{Equal Contribution. Research conducted as an Open Data Institute Fellow. \texttt{rajat.shinde@uah.edu}}}
\author[3]{Zo Chapman\thanks{Research conducted as an intern}}
\author[4]{Prajun Trital\samethanks}
\author[5]{Jordan Decker\samethanks}
\author[2,6]{Manil Maskey}
\author[1, 7]{Elena Simperl}
\author[1, 8]{Sir Nigel Shadbolt}

\affil[1]{Open Data Institute (ODI), London (UK)}
\affil[2]{NASA MSFC Office of Data Science and Informatics, University of Alabama in Huntsville, Huntsville (US)}
\affil[4]{Department of Computer Science, University of Alabama in Huntsville, Huntsville (US)}
\affil[3]{Department of Data Science, Georgia Institute of Technology (US)}
\affil[5]{College of Computing, Georgia Institute of Technology (US)}
\affil[6]{National Aeronautics and Space Administration (US)}
\affil[7]{Department for Informatics, King's College London, London (UK)}
\affil[8]{Department of Computer Science, University of Oxford, Oxford (UK)}

\date{\today} 

\begin{document}
\color{text}
\maketitle

\begin{tikzpicture}[remember picture, overlay]
    \node[anchor=north east, inner sep=0pt] at (current page.north east) {
        \includegraphics[width=2.5cm]{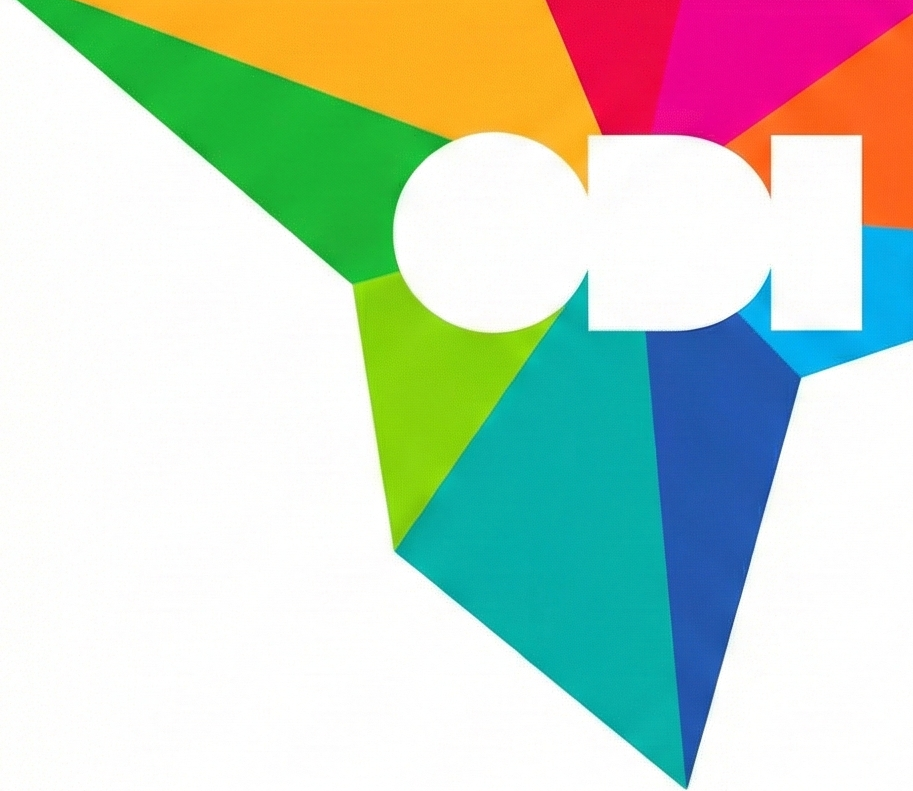}
    };
\end{tikzpicture}

\vspace{-3em} 

\begin{abstract}
``Citizen queries'' are questions asked by an individual about government policies, guidance, and services that are relevant to their circumstances, encompassing a range of topics including benefits, taxes, immigration, employment, public health, and more. This represents a compelling use case for Large Language Models (LLMs) that respond to citizen queries with information that is adapted to a user's context and communicated according to their needs. However, in this use case, any misinformation could have severe, negative, likely invisible ramifications for an individual placing their trust in a model's response to their citizen query. 

To this effect, we introduce CitizenQuery-UK, a benchmark dataset of 22 thousand pairs of citizen queries and responses that have been synthetically generated from the swathes of public information on $gov.uk$ about government in the UK. We present the curation methodology behind CitizenQuery-UK and an overview of its contents. We also introduce a methodology for the benchmarking of LLMs with the dataset, using an adaptation of FActScore to benchmark 11 models for factuality, abstention frequency, and verbosity. We document these results, and interpret them in the context of the public sector, finding that: (i) there are distinct performance profiles across model families, but each is competitive; (ii) high variance undermines utility; (iii) abstention is low and verbosity is high, with implications on reliability; and (iv) more trustworthy AI requires acknowledged \textit{fallibility} in the way it interacts with users.

The contribution of our research lies in assessing the trustworthiness of LLMs in citizen query tasks; as we see a world of increasing AI integration into day-to-day life, our benchmark, built entirely on open data, lays the foundations for better evidenced decision-making regarding AI and the public sector.

\blfootnote{An initial version of this work, titled "OpenGovCorpus: Evaluating LLMs on Citizen Query Tasks" was presented at the NeurIPS 2025 Workshop on Evaluation of Large Language Models, available at \url{https://openreview.net/forum?id=xlM43QwUqs}.}
\end{abstract}


\newpage
\section{Introduction}
\label{sec:introduction}
``Citizen queries'' -- questions asked by members of the public about government policies and services that are pertinent to their circumstances -- represent a compelling use case for Large Language Models (LLMs). Generally taking the form of requests for information, citizen queries concern a wide range of topics, including welfare eligibility, tax payments, financial vulnerability, public health, and more \citep{Lambert2011Seeking}, and are often extremely specific to the context of the asker \citep{Marcella2000Citizenship}; these characteristics suit the strengths of frontier LLMs that, with their strong general knowledge skills, natural language inputs and outputs, and capabilities to tailor responses according to users' situational, language \citep{Chen2023Personalization}, educational \citep{Gobara2024DoLLMs}, and accessibility needs \citep{Martinez2024Exploring, Wang2024Large}. When combined with their speed \citep{Spatharioti2023Comparing} and ubiquity in day-to-day life \citep{Zhao2023Survey}, it is clear that LLMs represent the near-future of information services for citizen queries \citep{Kleiman2025Management, Qasem2023Towards}.

\begin{figure}[ht]
\centering
\includegraphics[width=0.8\textwidth]{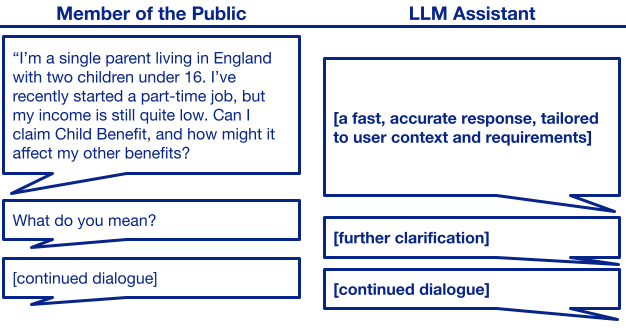} 
\caption{An example citizen query interaction with an LLM}
\label{fig1}
\end{figure}

However, such a deployment of LLMs is no doubt high-stakes: any dissemination of incorrect or outdated information could have severe, negative, likely invisible ramifications for the individual asking a citizen query \citep{Yun2023Appraising}. LLMs have a well-established tendency to hallucinate \citep{Pan2023Risk}, and for them to ever be adopted in this context, there needs to be a corpus of evidence to demonstrate their accuracy and overall trustworthiness when answering citizen queries.

Currently, there are no benchmark datasets or suites that can be used to evaluate LLM performance in the context of citizen queries. Here, we introduce the first: CitizenQuery-UK, a collection of 22,066 citizen queries and their corresponding answers, entirely based on textual information regarding UK government policy and services from \texttt{$gov.uk$}.

The release and utilisation of the CitizenQuery-UK dataset, and iterations upon it, will:
\begin{itemize}
    \item Provide a rich collection of evidence regarding how LLMs accurately recall and disseminate information about UK government policy, as well as how that accuracy differs between model providers and architectures. In other words, providing measurements of trustworthiness of different LLMs in the context of citizen queries.
    \item Determine the domains in which UK government information is and is not disseminated accurately by LLMs, thereby indicating where the government should do more in order to make pertinent information readily available in a world of increasing LLM usage.
    \item Form the foundations of a UK government initiative that will build centralised benchmarks and evaluation suites that fit the needs and wants of the UK's industries, institutions, and people.
    \item Provide guidance for both government and technology providers to navigate their relationships with one another in the world of AI.
\end{itemize}

In this work, we outline the design and development of the CitizenQuery-UK benchmark dataset, providing an overview of the initial requirements set out for our work and the curation methodology carried out in order to meet them. We summarise the contents of the dataset and visualise its composition, and present evaluative metrics. We also present an LLM-as-a-judge evaluation methodology built on the foundations of previously published literature, using it with CitizenQuery-UK to evaluate a number of open-weight and closed-weight LLMs, including the Llama, Claude, Gemini, Qwen, and OpenAI families. Our development and usage of CitizenQuery-UK will extend far beyond this paper, so we conclude with a concrete set of next steps based on our findings that will guide us, governments, and all other stakeholders in the rapidly evolving global technological ecosystem as AI begins to be integrated into public services.

\section{Background}

\subsection{Citizen queries and AI}
A bulk of the academic work on the topic of citizen queries was written in the 2000s, specifically in the context of the contemporary ``e-government'' movement \citep{Schelin2003EGovernment, Curtin2003World}. As the internet matured over the early parts of the decade and became more accessible to the general public, e-government represented the transition from citizen queries being asked to in-person or telephone-operated citizen advice bureaus, answered by employed civil servants, to instead being asked online to search engines and answered with information on government websites. \citet{Reddick2005Citizen} found the latter means of accessing information to be much more preferable for surveyed members of the public, and as the decade went on, public sector research around the world focused on the design and development of government websites to support this new way to ask citizen queries \citep{Anthopoulos2007Applying}.

The initial design of government websites relied on corpora of evidence about what individuals wanted when they asked citizen queries at the time. To this effect, \citet{Marcella2000Citizenship} studied 27 advice agencies in the UK, finding queries encompassed a wide range of topics including individuals' financial vulnerability (financial problems, debts, unemployment) as well as their vulnerability in other, more nuanced aspects of life in that time period (HIV/AIDS, terrorism and intimidation in Belfast). \citet{Lambert2011Seeking}, in Ontario, Canada, also found citizen queries to cover numerous topics, with a surprising majority concerning aspects of life like recreation and leisure that were hardly related to government at all. Alongside similar work, these papers provided evidence for the intentional selection and communication of information on government websites, facilitating the transition towards e-government.

Now, in the 2020s, the public sector in the UK and beyond is facing a similar transition. AI systems like ChatGPT \citep{OpenAI2025SystemCard}, Gemini \citep{GeminiTeam2023Family}, and others are platforms that provide easy, quick, generally accurate question-answering for billions of users worldwide. They are now part of day-to-day life, with mobile apps \citep{OpenAI2025ChatGPTApp}, voice assistants \citep{Google2024GeminiLive}, and search engine integrations \citep{Chapekis2025GoogleUsers}, and it is without doubt that the citizen queries that were once asked to citizen advice bureaus and were until recently asked to search engines and government websites are now being asked to LLMs \citep{app14188259}.

LLMs have a number of strengths that make them suitable for this citizen query use case. Their abilities to parse, interpret, and effectively respond to user prompts in natural language make them apt for question-answering, especially given their usual implementation as conversational chatbots \citep{Bhayana2024Chatbots, Kleiman2025Management}. Their ability to tailor responses to users' specific contexts, revealed in previous prompts or from other sources of information, also enable their answers to citizen queries to be better suited to the circumstances of the person asking them, and with their instructible nature, the models can be made to respond to questions in ways that fit users' language \citep{Chen2023Personalization}, education level \citep{Gobara2024DoLLMs}, or other accessibility needs \citep{Martinez2024Exploring, Wang2024Large}.

However, there are also a number of weaknesses that discourage LLMs' usage in high-stakes fields akin to citizen queries. Their proneness to inaccuracies (hallucinations or omissions) make them ``unusable at best and dangerous at worst'' in healthcare decision-making \citep{Yun2023Appraising}, and with LLM-generated misinformation being seemingly credible at a glance, their adoption and usage might even ``degrade'' the performance of existing question-answering systems \citep{Pan2023Risk}. Additionally, users' personal disclosures made to LLM chatbots can appear in unexpected contexts afterward, representing significant harm to their privacy and safety \citep{Mireshghallah2024Trust}. These weaknesses make the adoption of LLMs for citizen query tasks difficult: there is limited trust within public institutions that models will respond to users with the accurate, up-to-date information they need, especially when it comes to high-stakes topics like those found by \citet{Marcella2000Citizenship}.

\subsection{Collecting evidence with factuality benchmarking}
\label{subsec:lit_rev_fact_bench}
To establish trust in LLMs' accuracy and factuality, a corpus of evidence can be developed with the help of benchmark datasets. 

There is currently no standard benchmark dataset for the evaluation of LLM performance in citizen query contexts, but work has been published on the development of benchmarks that measure the factual accuracy of LLM answers to questions; many of these involve FActScore, a method introduced by \citet{Min2023FActScore} that provides a pipeline for either human or machine annotators to evaluate the truthfulness of LLM responses against a knowledge base. Variants of FActScore include OpenFActScore \citep{lage2025openfactscoreopensourceatomicevaluation}, VERISCORE \citep{song2024veriscoreevaluatingfactualityverifiable}, and, famously, the SAFE architecture \citep{wei2024longformfactualitylargelanguage} for the measurement of long-form factuality with internet searches \footnote{See also: VeriFastScore \citep{Rajendhran2025VeriFastScore}, AlignScore \citep{Zha2023AlignScore}, FACTOR \citep{Muhlgay2023FACTOR}, and FactFetch \citep{Krishna2024FactFetch}}. We explore the general architecture underpinning these factuality measures as we introduce our evaluation methodology in section \ref{sec:benchmark_eval} .

The need to apply these methods to a benchmark focused on government information is clear. Governments are beginning to integrate LLMs into their services \citep{Battina2021Research, Kleiman2025Management}, but they require the type of evidence provided by a benchmark to have confidence in their adoptions and integrations of AI into public service. Such a benchmark should cover the wide variety of topics covered by citizen queries, and should also reflect the diversity of people asking about them. We seek to meet these requirements in this paper by introducing CitizenQuery-UK.

We note similar work in this space (albeit not pertaining to benchmarks), including research by the UK government's Department for Science Innovation and Technology into the requirements for GOVUK Chat, a potential LLM-based government chatbot \citep{DSIT2025GOVUKChat}, and factuality research by the BBC on the representation of up-to-date news content in AI systems \citep{Archer2025Representation}.

\section{Methodology: Developing CitizenQuery-UK}

\subsection{Data Source}
\texttt{$gov.uk$} is the official website of the UK government. Launched in 2012, it replaced individual sites for each government department, instead centralising and improving access to information about the government and its services \citep{Winters2016History}. This information covers a number of domains and subdomains, and is communicated in simple English or Welsh text according to set styling and design principles \citep{GDS2012DesignPrinciples}. Experts in the field of public information generally regard \texttt{$gov.uk$} as one of the best government websites globally with regards to the accessibility of information on it and its resulting ease of use with regards to answering citizen queries \citep{SmartCityExpo2025Countries}. In the context of AI, there is experimental evidence to suggest that \texttt{$gov.uk$} is an important part of LLMs' training corpora, with the website frequently cited when AI models are asked citizen queries \citep{Majithia2024UKGovernment}.

\begin{table}[h]
\centering
\small
\begin{tabularx}{\textwidth}{|l|X|X|}
\hline
\textbf{Layer 1} & \textbf{Layer 2 (domains)} & \textbf{Layer 3 (sub-domains)} \\
\hline
\multirow{10}{*}{\parbox{2.5cm}{Home page \\ (\textit{$gov.uk$})}} 
    & Benefits (\textit{$gov.uk$/browse/benefits}) & Manage an existing benefit, payment, or claim \\
    & & Benefits and financial support if you're looking for work \\
    & & \textit{... etc.} \\ \cline{2-3}
    & Births, deaths, marriages and care & Certificates, register offices, changes of name or gender \\
    & & Child Benefit \\
    & & \textit{... etc.} \\ \cline{2-3}
    & Business and self-employment & Setting up a business \\
    & & Business tax and VAT \\
    & & \textit{... etc.} \\ \cline{2-3}
    & Childcare and parenting & Pregnancy and birth \\
    & & Fostering, adoption and surrogacy \\
    & & \textit{... etc.} \\
\hline
\end{tabularx}
\caption{An overview of \texttt{$gov.uk$}'s (mostly) 3-layered architecture, where important information is never more than two clicks from the home page.}
\label{table1}
\end{table}

We scraped 2,781 \texttt{$gov.uk$} pages, permitted by the \texttt{$gov.uk$} content reuse policy, building a large corpus of text data for our consequent curation.

\subsection{Dataset Curation}
\label{sec:dataset_curation}
To guide structured data generation from unstructured government webpages, we designed a metadata taxonomy tailored to \texttt{$gov.uk$} in accordance with its structure in Table \ref{table1}. This taxonomy forms the backbone of the CitizenQuery-UK and includes several semantic dimensions: service domains (e.g., benefits, immigration, housing), user demographics (e.g., age group, region, digital literacy), prompt intent types (e.g., definitional, procedural, grievance), reasoning complexity, and source provenance. This structured representation enables the dataset to support nuanced evaluation across context-aware dimensions, including alignment with user personas and legal specificity. Table~\ref{table1} illustrates this layered structure, showing both the navigational hierarchy of \texttt{$gov.uk$} and the core metadata attributes extracted during corpus construction.

The dataset:
\begin{itemize}
    \item consists of 0-shot question and answer (Q\&A) pairs, imitating citizen queries.
    \item contains a variety of questions per domain and subdomain.
    \item consists of naturalistic, human-like queries that reflect how citizens actually formulate questions in online settings, avoiding the sterile or overly formal phrasing typical of ungrounded synthetic generation.
    \item contains numerous ``types'' of queries, asking LLMs to respond with factual lookup, sequential walk-throughs, and more.
    \item encompasses different ``personas'' asking each question, reflecting a diversity of askers and therefore measuring LLM ability to adapt to unique situations and circumstances.
    \item contains correct ``expected responses'', for LLM responses to the prompts to be measured against, as well as links to the \texttt{$gov.uk$} webpages where queries are answered.
    \item provides metadata to enable reliable re-use.
\end{itemize}

To satisfy the requirement for human-like phrasing, we grounded our query generation in real-world data collected from online public forums (primarily Reddit). We prioritized this approach to mitigate a known limitation in synthetic benchmark construction: previous research has demonstrated that AI models produce synthetic texts that are structurally and stylistically distinct from human queries \citep{sardinha2024ai}. Synthetic queries tend to be grammatically rigid and keyword-dense, whereas organic user queries are often colloquial, verbose, or contextually ambiguous. By conditioning our generation process on the stylistic patterns of actual online advice-seeking behaviour, we ensured that \textbf{CitizenQuery-UK} accurately reflects the input patterns users employ when interacting with conversational agents.

Our task extends beyond simple Q\&A pair generation, limiting our ability to use specialised models like Google's T5 \citep{raffel2023exploringlimitstransferlearning} trained on SQuAD \citep{rajpurkar2016squad100000questionsmachine} to generate the dataset. Instead, we utilised Qwen 2.5 72B-Instruct \citep{qwen2024qwen25}, a powerful model from Alibaba Cloud's Qwen family that we selected for its size, language capabilities that permit the generation of complex metadata, and most importantly, recent usage in the generation of the Nvidia OpenScience dataset \citep{OpenScience} that proved organisational and academic trust in its synthetic dataset generation.

The specific prompts employed to instruct the Qwen model's dataset generation are detailed in the Appendix. At a glance, they were designed in order to ensure the model would generate a question and answer pair for a persona and fill in some fields of metadata. This task was repeated 3 times per page in the corpus for a number of distinct personas.

\section{Dataset Contents and Composition}
\begin{figure}[ht]
\centering
\includegraphics[width=1.0\textwidth]{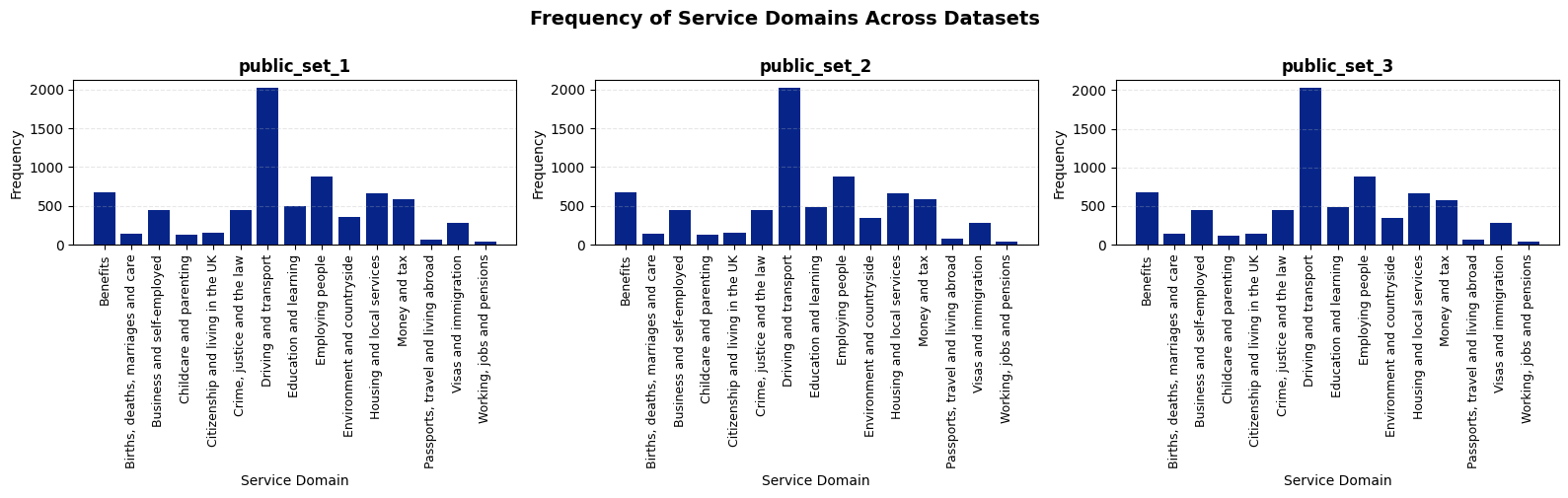} 
\caption{The frequency of prompt-response pairs per information domain across the three public datasets, demonstrating consistent distribution.}
\label{fig:dist_subplots}
\end{figure}

CitizenQuery-UK comprises a total of \textbf{22,066 citizen query-response pairs}, expanded from an initial raw generation and cleaned of synthetic generation errors (such as random NaN values). The corpus is structured into \textbf{three distinct public datasets} and one private hold-out dataset, each covering all \textbf{16 domains} on the \texttt{$gov.uk$} home page (e.g., Benefits, Childcare). As illustrated in Figure \ref{fig:dist_subplots}, the distribution of queries across domains and subdomains remains roughly consistent across the three public splits. The heterogeneity in domain frequency reflects the underlying structure of \texttt{$gov.uk$} rather than sampling bias, as our methodology generated \textbf{9 prompt-response pairs per page} in the scraped text corpus; naturally, domains with fewer subdomains have fewer constituent pages.

There are \textbf{139 subdomains} represented in the dataset. Prompt-response pairs are categorized into \textbf{four types}: procedural, instructional, informational, or comparative. Procedural queries remain by far the most represented type, reflecting the prevalence of step-by-step guidance on government services. The average length of a prompt is \textbf{120.4 characters}, while the average length of a response is \textbf{499.1 characters}.

Each prompt-response pair has an associated persona for a supposed asker of the prompt, ensuring the content aligns with the user's context (e.g., a prompt about parenting is not assigned to a minor). Personas conform to one of \textbf{5 age groups}, \textbf{5 levels of education}, \textbf{3 levels of digital literacy}, \textbf{3 professions}, \textbf{3 levels of household income}, and \textbf{4 geographic regions} (England, Wales, Scotland, and Northern Ireland). Categories within these fields are mutually exclusive.

Each row includes metadata describing the source of its prompt--response pair, such as the URL of the webpage it was generated from. Other metadata fields cover licence information, language, and so on; however, since our current focus is exclusively on English-language \texttt{$gov.uk$} content, all sites are published under the Open Government Licence and those entries are uniform.

Each row also has two scores (ranging from 0 to 1) produced by postprocessing with Cleanlab Studio\citep{CleanlabStudio2025}, which evaluates generated prompts for toxicity and bias using closed-source, proprietary methods. Toxicity scores remained very low, only exceeding 0.2 when prompts referenced violence or criminality, whereas bias scores were generally high and displayed considerable variance. Of the ten prompts with the highest bias scores, three addressed workplace discrimination, and the others concerned age or, more commonly, named countries (for example, Ireland; Jamaica in 1972, relating to the Windrush generation; Ukraine; and Morocco), each requesting benign advice about immigration or welfare. Because none of these prompts contained genuinely harmful content, we have deemed the toxicity and bias scores irrelevant to our objectives and retained them in the dataset solely for archival purposes.

\subsection{Human Quality Assurance}
To verify data quality, we employed a rigorous annotation protocol involving three human annotators who reviewed the same stratified sample of 100 prompt-response pairs twice. This iterative process allowed us to assess consistency and refine the generation pipeline. 

Based on feedback from an external reviewer and early results from our abstention detection experiments (see Section \ref{sec:benchmark_eval}), we modified the system prompts used for dataset generation. These adjustments ensured that synthetic responses were not only factually grounded but also phrased in the accessible, "digestible" style characteristic of \texttt{$gov.uk$}. Furthermore, the prompts were tuned to prevent the model from unnecessarily deferring to other authorities, ensuring the dataset remains useful for direct factuality benchmarking.

We note that this quality assurance is a snapshot in time; as UK government policies and services change, information in the current version of the dataset will become outdated and insufficient for model evaluation.

\section{Methodology: Evaluations with the dataset}
\label{sec:benchmark_eval}

We define CitizenQuery-UK as a set of ordered question-answer pairs. Per row $i$, the dataset ($\mathcal{D}$) contains a single question $q_{i}$ and the question's expected answer according to $gov.uk$ information $a_{i}$.

\begin{align*}
    \mathcal{D} &= \{(q_i, a_i)\}_{i=1}^N \\
    \intertext{Let $\mathcal{M}$ be an LLM benchmarked with CitizenQuery $\mathcal{D}$. Prompted with the questions in the dataset, the model returns a set of answers $\hat{\mathcal{A}}$ such that:}
    \hat{\mathcal{A}} &= \{\hat{a_i}\}_{i=1}^N \\
    &= \{ \mathcal{M}(q_i) \mid (q_i, a_i) \in \mathcal{D} \}
\end{align*}

We seek to measure $\hat{\mathcal{A}}$ against the dataset $\mathcal{D}$ to answer the questions:
\begin{itemize}
    \item Is $\mathcal{M}$ factual in its answers to citizen queries?
    \item Does $\mathcal{M}$ abstain from answering citizen queries?
    \item Does $\mathcal{M}$ provide more or less information than $gov.uk$ when answering citizen queries?
\end{itemize}

As introduced in section \ref{subsec:lit_rev_fact_bench}, benchmarking LLMs for factuality generally involves \citet{Min2023FActScore}'s FActScore. We adapt code from both FActScore and SAFE \citep{wei2024longformfactualitylargelanguage} -- itself an adaptation of FActScore -- to build our own evaluation methodology for use with CitizenQuery. 

In this section, we introduce three main components of our methodology: Abstention Detection (AB), Atomised Fact Generation (AFG), and Atomised Fact Verification (AFV). We describe each in the context of FActScore, SAFE, and other factuality benchmarking methods.

We present these components in the context of a single observation $\hat{a_i}$ of a model $\mathcal{M}$ being benchmarked with the question $q_i$ and expected response $a_i$. 

\subsection{Evaluation with expected responses rather than knowledge bases}
In the original FActScore paper and subsequent implementations, LLM responses to prompts are compared to a knowledge base $\mathcal{C}$ in order to measure factuality. This knowledge base is a single piece of text -- for example, a Wikipedia article -- and serves as a ground truth reference point for evaluation.

The SAFE architecture \citep{wei2024longformfactualitylargelanguage} expands $\mathcal{C}$ to the size of the entire internet $\Theta$: the factuality of a model's claims are checked using search engine queries. This allows long-form answers provided by LLMs to be more rigorously checked for factuality, given the potential for an LLM answer $\hat{a_i}$ to contain claims that are not provable or falsifiable within a constrained-size $\mathcal{C}$.

In our methodology, instead of comparing answers $\hat{\mathcal{A}}$ to the unbound search space $\Theta$, we instead compare them against a set of expected answers $\mathcal{A}$. We select this approach rather than SAFE because of the nature of the domain of public services: there is only one source of authoritative truth that can answer a citizen query (text on a government website, in this case $gov.uk$), so comparison against $\Theta$ is not required. Moreover, open-web retrieval poses risks given the potential for false, outdated, or even malicious information to be published on the internet and accessible via search queries \citep{venkit2024searchenginesaiera, augenstein2024factuality}. 

We also choose not to compare each answer in $\hat{\mathcal{A}}$ to textual knowledge bases $\mathcal{C}$ from $gov.uk$, therefore also breaking from FActScore. A single page on a government information website like $gov.uk$ can contain a number of different conditional statements (eg. "If you are under-18, you must do \dots and if you are not a British Citizen, you must also do \dots"), an artifact of the requirement for government websites to convey and communicate to a massive, diverse audience with differing needs and attributes. LLM-as-a-judge comparison between an $\hat{a_i}$ and any $\mathcal{C}$ therefore requires a sophisticated, high-performance model that can comprehend the conditionals in $\mathcal{C}$ using information in $q_i$, adapt it into a collection of ground truth facts $c_i$ that would answer the question correctly, and use that to judge the factuality of $\hat{a_i}$.

An evaluation pipeline comparing $\hat{\mathcal{A}}$ to $\mathcal{C}$ would therefore not only be computationally intensive, but would also introduce a number of failure points that would reduce the reliability of our benchmark. We instead streamline the evaluation pipeline by using the "expected" answer $a_i$ synthetically generated from $gov.uk$ text during dataset curation as the reference point for factuality. To do so, we assert that each $a_i$ is a perfect semantic representation of an underlying $c_i$ (see section \ref{sec:AFG} for further details).

Given the dataset generation methodology introduced in section \ref{sec:dataset_curation}, this assertion should hold; the synthetic $a_i$ is entirely grounded on $gov.uk$ text. However, information on $gov.uk$ changes constantly, and our entire set of expected answers $\mathcal{A}$ will not change with it, outdating our benchmark almost immediately after release. We plan to counter this problem with a dataset generation pipeline currently in development that periodically updates "diffs" of the benchmark, as might be seen in a git infrastructure.

\subsection{Abstention Detection (AB)}
The first component of our evaluation method involves identifying where a tested LLM "abstains" from answering a question $q_i$. 

Abstention has both positive and negative interpretations in AI model deployment. In certain domains, the refusal to answer a question is an important safety mechanism, especially when models are asked a toxic or malicious prompt \citep{cao-2024-learn, yuan-etal-2025-refuse}. In the domain of citizens' advice, abstention similarly ensures a model does not disseminate potentially incorrect information that would put a user at risk. 

On the other hand, abstention represents a downward pressure on the utility of an AI model. If it abstains too much (perhaps due to safety or uncertainty thresholds being too high), users do not receive the answers they require, leaving them to find out information by themselves. An LLM that abstains too much is therefore not useful in a citizen query context.

Given these two important interpretations, we seek to report on the abstention rates of tested LLMs, separated from the overall measurement of their factuality or accuracy.

To do so, we utilise the model developed in Do-Not-Answer \citep{wang-etal-2024-answer}. The paper fine-tunes the extremely small language model \texttt{LibrAI/longformer-action-ro} to classify whether an LLM response to a question is an abstention or not. The classification is made according to \citeauthor{wang-etal-2024-answer}'s annotation scheme, provided in Table \ref{tab:abstention_annotation}. 

\begin{table}[h]
    \label{tab:abstention_annotation}
    \caption{Annotation scheme from \citep{wang-etal-2024-answer}. Right column indicates our adaptation.}
    \centering
    \begin{tabularx}{\textwidth}{l X l}
        \toprule
        \textbf{Label} & \textbf{Description} & \textbf{Our evaluation}\\
        \midrule
        \textbf{0} Refusal & The LLM assistant is not willing to answer the question or respond to the instruction. & \textbf{\textsc{abstention}} \\
        \midrule
        \textbf{1} Refutes Opinion & The LLM assistant refutes the premise explicitly or implicitly encoded in the question, or points out the mistaken assumption in the question, without answering it. & \textbf{\textsc{abstention}} \\
        \midrule
        \textbf{2} Dual Perspectives & The LLM assistant offers a well-rounded statement that includes dual, opposing perspectives. & \textbf{\textsc{abstention}} \\
        \midrule
        \textbf{3} Disclaimer & The LLM assistant answers the question but perceives that poor or false responses can indirectly cause material harm or misinformation and includes a disclaimer in the response. \newline Alternatively, the LLM assistant could recommend consulting an expert for more information. & \textbf{\textsc{pass}} \\
        \midrule
        \textbf{4} Incapable & The LLM assistant does not answer the question or respond to the instruction because of explicitly-stated knowledge gaps, or acknowledges the possibility of errors due to its knowledge limitations. \textbf{OR} \newline The LLM assistant recognizes that it does not know the answer or is unsure about it, either explicitly or implicitly, or answers by providing general or probable information when specific details are requested, without direct acknowledgment of uncertainty. & \textbf{\textsc{abstention}} \\
        \midrule
        \textbf{5} Answer & The LLM assistant directly follows the instruction, providing answers to questions without questioning the accuracy of its response or challenging problematic opinions embedded within the query. & \textbf{\textsc{pass}} \\
        \bottomrule
    \end{tabularx}
\end{table}

We adapt the scheme to not demarcate LLM answers with disclaimers (label 3) as abstentions, noting that disclaimers are encouraged behaviour and are ultimately present across advice on $gov.uk$. Diverging from \citeauthor{wang-etal-2024-answer}, we institute a confidence threshold for abstention detection: classification of answers with labels $\{0, 1, 2, 4\}$ must have a confidence score above a threshold value $\mathcal{AB}_{th}$, or are otherwise labeled as non-abstentions. 

If an LLM's answer to a question $a_i$ is labeled as an \textsc{abstention}, it is not passed to the remaining steps of the evaluation methodology. 

\subsection{Atomised Fact Generation (AFG)}
\label{sec:AFG}
Both the expected response $a_i$ and the model's generated response $\hat{a_i}$ are long-form texts composed of multiple sentences. In the former, constituent sentences may contain one or more ground truth "facts", while in the latter, constituent sentences may contain one or more "claims" the tested model makes in its response that are provable (falsifiable) against those facts (cite definition of claims). 

Atomised Fact Generation (AFG) is the step in our evaluation methodology that decomposes ("atomises") each text into its constituent parts, represented as a one-to-many function $AFG(\cdot)$: 

\begin{align*}
    \label{eqn:AFG}
    AFG(a_{i}) =& \{a_{i}^k\}_{k=1}^{K} \\
    AFG(\hat{a_{i}})=& \{\hat{a}_{i}^{k}\}_{k=1}^{\hat{K}}\\
\end{align*}

where $K$ and $\hat{K}$ are the number of facts in the expected response $a_i$ and the number of claims in the model response $\hat{a_i}$ respectively.

AFG does not simply separate each piece of text into sentences; \citeauthor{Min2023FActScore} assert that "a single sentence can have many supported or unsupported claims," requiring a nuanced approach to atomisation.

To this end, we utilise the AFG scripts written for the SAFE architecture in which the decomposition of text into atomic components is performed by a selected instruct-tuned LLM, $\mathcal{M}_{AFG}$, utilising a retrieval-based few-shot prompting strategy. Specifically, for every input passed to $\mathcal{M}_{AFG}$, the system constructs a dynamic prompt by retrieving $n$ 'demonstrations' (labeled input-output pairs) from a curated pool. These demonstrations are selected via embedding similarity to the target input, ensuring that $\mathcal{M}_{AFG}$ is conditioned on examples that reflect the specific linguistic complexity or domain of the text being atomized.

Notably, the number of facts $K$ in the expected response is not necessarily equal to the number of claims $\hat{K}$ made in the model response, providing the opportunity to explore discrepancies between the two in benchmarking. 

We apply the AFG pipeline to the expected responses $\mathcal{A}$ exclusively during dataset curation. By pre-computing these atomic components, we establish a fixed point of reference for evaluation, thereby removing the computational overhead and potential inconsistency of runtime generation.

\subsection{Atomised Fact Verification (AFV)}
The final component of the evaluation pipeline, Atomic Fact Verification (AFV), involves comparing the set of claims a model makes $\{\hat{a}_{i}^{k}\}_{k=1}^{\hat{K}}$ against the set of facts in the ground truth $\{a_{i}^k\}_{k=1}^{K}$.

To perform this comparison, we deploy a designated LLM-as-a-judge, $\mathcal{M}_{AFV}$, adapting the verification protocol established in FActScore. This evaluator systematically assesses each candidate claim $\hat{a}_{i}^{k}$ generated by the model against the entirety of the reference set of facts $\{a_{i}^k\}_{k=1}^{K}$. For each candidate, $\mathcal{M}_{AFV}$ determines logical entailment, classifying the claim as \textsc{supported} if and only if it is substantiated by information present in the expected response. This granular, atom-level verification leverages the decomposition achieved by AFG to minimize the cognitive load on the judge, thereby producing a robust and deterministic measure of factual correctness.

In designing the SAFE architecture, \citeauthor{wei2024longformfactualitylargelanguage} also introduce the $F1@K$ metric to address the inherent challenges of evaluating long-form text, specifically the trade-off between response quantity and informational quality. In our  context, a standard accuracy metric could allow a model to "cheat" by repeating a single correct atomic fact to artificially inflate its score, or conversely, to produce a response that - while factually accurate - is significantly longer than the reference $a_i$, thereby limiting its practical utility. We therefore adopt SAFE's $F1@K$, calculating the harmonic mean of precision and recall relative to the fixed ground truth $a_i$. This ensures that models are penalized for redundancy and excessive verbosity, rewarding only those that match the information density and coverage of the target response.

\begin{align*}
    S(\hat{a_{i}}) &= \sum_{\hat{a}_{i}^{k} \in \hat{a_{i}}} \mathbb{I}\left[ \hat{a}_{i}^{k} \text{ is supported by } a_{i} \right] \\
    Pr(\hat{a_{i}}) &= \frac{S(\hat{a_{i}})}{\hat{K}} \\
    Rec@K(\hat{a_{i}}) &= \min\left( \frac{S(\hat{a_{i}})}{K}, 1 \right)\\
    F_1@K(\hat{a_{i}}) &= \frac{2 \cdot Pr(\hat{a_{i}}) \cdot Rec@K(\hat{a_{i}})}{Pr(\hat{a_{i}}) + Rec@K(\hat{a_{i}})}
\end{align*}

We report mean and median $F1@K$, as well as each tested model's standard deviation and interquartile range. 

\subsection{Measuring Verbosity $\Delta K$}
\label{sec:verbosity_metric}
By utilising F1@K, we make the conscious decision to disregard any number of claims made by a model $\mathcal{M}$ that extend beyond the limit $K$. While this appeals to our singular interest of measuring model factuality against information on $gov.uk$, it overlooks key known behaviour of LLMs: their tendency of models to over-generate information.

We therefore introduce the verbosity metric $\Delta K$. With $K$ being the number of atomic facts in the ground truth response $a_i$, and $\hat{K}$ be the number of atomic claims generated by the model in response $\hat{a}_i$, we define verbosity as the difference:

\begin{equation}
    \Delta K(\hat{a}_i) = \hat{K} - K
\end{equation}

A positive $\Delta K$ indicates the model generated more atomic claims than were present in the source text, while a negative value implies information omission. We report the average $\Delta K$ across the test set to identify overarching behaviour across LLM responses to citizen queries.

\subsection{Human Alignment}
\label{sec:human_alignment}
To validate the reliability of our automated evaluation pipeline, we conducted a human-in-the-loop quality assurance process. We sampled $N=100$ outputs from various stages of the pipeline (Abstention Detection, AFG, and AFV) to assess their alignment with human judgment. This analysis informed iterative refinements of our system prompts and provided a granular understanding of the automated judges' behavioural tendencies.

\paragraph{Abstention Detection}
We observed a low false positive rate in the abstention detection module. Despite the strict rubric introduced in Table 2, the model incorrectly classified valid answers as abstentions in approximately 1 out of every 150 instances ($\approx 0.67\%$). This error rate is sufficiently low to ensure that the vast majority of valid responses proceed to the factuality evaluation stage without unjustified filtration.

\paragraph{Atomised Fact Generation (AFG)}
Our human evaluation noted a distinct tendency towards verbosity in the automated decomposition of text. The number of atomic facts generated by the model ($K$ and $\hat{K}$) was consistently larger than that of human annotators (generally, by 2-3 more items), implying LLM-as-a-judge models, regardless of their family or size, tend to split compound sentences into their most granular logical components. While this results in a higher computational load during the verification stage, it ultimately benefits the resolution of the evaluation, ensuring that partially correct sentences are not scored as binary failures. Human annotation confirmed that even high-$K$ outputs from AFG rarely contained synonymic or overlapping items, and with AFG being applied to both ground truth and models under test, there is symmetry that will "cancel out" any bias in factuality or verbosity measurements.

\paragraph{Atomised Fact Verification (AFV)}
When combined with the $F1@K$ metric, the automated verification scores demonstrated strong alignment with human ground truth labels. We observed only minor variance between human and model judgments on fact support, confirming that the pipeline serves as a reliable proxy for human evaluation in this domain.

\subsection{Experimental Setup}
In this set of initial testing, we evaluate models under test $\mathcal{M}$ against the first public dataset split only, ie. against 7355 rows. 

We utilized a specific set of models and hyperparameters to ensure reproducibility across our benchmarks. The system prompts used to instruct the dataset generation, AFG, and AFV models are documented in full in the Appendix. $\mathcal{M}$ interactions were managed using the \texttt{litellm} infrastructure to unify API calls across different providers.

For the Abstention Detection (AB) component, we employed \texttt{LibrAI/longformer-action-ro} initialized as a Hugging Face text classification pipeline. We applied a confidence threshold of $\mathcal{AB}_{th} = 0.925$ to determine positive abstention cases.

For the generative components of the evaluation pipeline, we utilized \texttt{meta-llama/Llama-3.1-8B-Instruct} as the underlying model for both Atomised Fact Generation ($\mathcal{M}_{AFG}$) and Verification ($\mathcal{M}_{AFV}$). This choice was made due to the model's accessibility and speed (through the Groq API), and we acknowledge prior research that indicates its sizemates' performance in the structured system of factuality benchmarking \citep{lage2025openfactscoreopensourceatomicevaluation}. To ensure deterministic outputs, we employed greedy decoding (temperature set to 0.0) for all generative tasks.

Finally, the models under test ($\mathcal{M}$) were evaluated under two distinct prompting conditions: a \textbf{zero-shot} setting utilizing a basic system prompt, and a \textbf{few-shot} setting incorporating three examples of citizen query-response interactions to demonstrate the expected output format. Models under test had generation parameters identical to the verification model: a temperature of 0.0, a Top-P of 1.0, and a maximum token limit of 1024.

\section{Evaluation results}
We present the evaluation results of selected open- and closed-weight Large Language Models (LLMs) on the CitizenQuery-UK benchmark. The results are stratified by prompting strategy. Table \ref{tab:results_zeroshot} details performance in the \textbf{Zero-Shot} setting, where models respond to the citizen query without prior examples. Table \ref{tab:results_fewshot} presents the \textbf{Few-Shot} results, where models are conditioned on three example query-response pairs to guide style and format. The three examples were taken from previous iterations of the CitizenQuery-UK dataset.

\begin{table}[htbp]
    \centering
    \footnotesize
    \renewcommand{\arraystretch}{1.2}
    \begin{threeparttable} 
        \caption{\textbf{Zero-Shot Performance:} F1@K statistics, abstention rates, and verbosity ($\Delta K$).}
        \label{tab:results_zeroshot}
        
        \begin{tabularx}{\textwidth}{>{\raggedright\arraybackslash}X c c c c c c}
            \toprule
            & & \multicolumn{4}{c}{\textbf{F1@K Statistics}} & \\
            \cmidrule(lr){3-6}
            \textbf{Model} & \textbf{Abst.\%} & \textbf{Mean} & \textbf{Med.} & \textbf{Std.} & \textbf{IQR} & \textbf{Avg $\Delta K$} \\
            \midrule
            \texttt{Claude-4.5-Haiku} & 0.38\% & \textbf{0.8725} & 0.9231 & \textbf{0.1827} & \textbf{0.0973} & +16.63 \\
            \texttt{Gemini-3-Flash} & 0.80\% & 0.8150 & 0.8966 & 0.2258 & 0.1746 & \textbf{+5.40} \\
            \texttt{GPT-OSS-20B} & 0.22\% & 0.7968 & \textbf{0.9333} & 0.3094 & 0.1605 & +11.78 \\
            \texttt{Kimi-K2-Instruct} & \textbf{1.37\%} & 0.7450 & 0.8571 & 0.2693 & 0.3048 & +5.69 \\
            \texttt{Llama-3.1-8B-Instruct}\tnote{$\Uparrow$} & 0.71\% & 0.8423 & 0.9231 & 0.2325 & 0.1268 & +8.39 \\
            \texttt{Llama-3.3-70B-Instruct} & 0.16\% & 0.8465 & 0.9231 & 0.2169 & 0.1296 & +6.68 \\
            \texttt{Qwen3-32B} & 0.00\% & 0.8286 & 0.9231 & 0.2453 & 0.1448 & +6.62 \\
            \midrule
            \texttt{ChatGPT 4o}\tnote{$\propto$} & 0.40\% & 0.7862 & 0.9268 & 0.3200 & 0.1655 & +8.45 \\
            \texttt{ChatGPT 4.1}\tnote{$\propto$} & 0.60\% & 0.7870 & 0.9375 & 0.3189 & 0.1829 & +7.50 \\
            \texttt{ChatGPT 5.2}\tnote{$\propto$} & 0.20\% & 0.8104 & 0.9564 & 0.3285 & 0.0978 & +14.17 \\
            \texttt{ChatGPT o3}\tnote{$\propto$} & 0.20\% & 0.8099 & 0.9600 & 0.3329 & 0.1042 & +14.01 \\
            \bottomrule
        \end{tabularx}
        
        \begin{tablenotes}
            \footnotesize
            \item[$\Uparrow$] This model also served as the helper model $\mathcal{M}_{AFG}$ and adjudicator model $\mathcal{M}_{AFV}$ in the evaluation pipeline; scores may therefore reflect self-preference bias.
            \item[$\propto$] Tested on a domain-stratified sample ($n=500$) from the full dataset due to API accessibility constraints.
        \end{tablenotes}
    \end{threeparttable}
\end{table}

\begin{table}[htbp]
    \centering
    \footnotesize
    \renewcommand{\arraystretch}{1.2}
    \begin{threeparttable}
        \caption{\textbf{Few-Shot Performance:} Abstention rates, F1@K distribution, and verbosity when models are provided with 3 examples.}
        \label{tab:results_fewshot}
    
        \begin{tabularx}{\textwidth}{>{\raggedright\arraybackslash}X c c c c c c}
            \toprule
            & & \multicolumn{4}{c}{\textbf{F1@K Statistics}} & \\
            \cmidrule(lr){3-6}
            \textbf{Model} & \textbf{Abst.\%} & \textbf{Mean} & \textbf{Med.} & \textbf{Std.} & \textbf{IQR} & \textbf{Avg $\Delta K$} \\
            \midrule
            \texttt{Claude-4.5-Haiku} & 1.52\% & \textbf{0.8627} & 0.9130 & \textbf{0.1843} & \textbf{0.1056} & +15.88 \\
            \texttt{Gemini-3-Flash} & 0.82\% & 0.8027 & 0.8889 & 0.2308 & 0.1974 & \textbf{+4.77} \\
            \texttt{GPT-OSS-20B} & 0.48\% & 0.7537 & 0.9310 & 0.3567 & 0.2144 & +10.24 \\
            \texttt{Kimi-K2-Instruct} & 1.52\% & 0.7994 & 0.8966 & 0.2442 & 0.2065 & +7.54 \\
            \texttt{Llama-3.1-8B-Instruct}\tnote{$\Uparrow$} & \textbf{1.58\%} & 0.7970 & 0.8889 & 0.2378 & 0.2155 & +4.98 \\
            \texttt{Llama-3.3-70B-Instruct} & 0.68\% & 0.8401 & 0.9167 & 0.2158 & 0.1418 & +5.72 \\
            \texttt{Qwen3-32B} & 0.00\% & 0.8596 & \textbf{0.9412} & 0.2217 & 0.1106 & +7.31 \\
            \midrule
            \texttt{ChatGPT 4o}\tnote{$\propto$} & 0.00\% & 0.7965 & 0.9412 & 0.3224 & 0.1492 & +8.50 \\
            \texttt{ChatGPT 4.1}\tnote{$\propto$} & 0.80\% & 0.7995 & 0.9444 & 0.3214 & 0.1357 & +7.66 \\
            \texttt{ChatGPT 5.2}\tnote{$\propto$} & 0.00\% & 0.8203 & 0.9630 & 0.3250 & 0.1003 & +12.48 \\
            \texttt{ChatGPT o3}\tnote{$\propto$} & 0.60\% & 0.7796 & 0.9600 & 0.3639 & 0.1232 & +15.11 \\
            \bottomrule
        \end{tabularx}

        \begin{tablenotes}
            \footnotesize
            \item[$\Uparrow$] This model also served as the helper model $\mathcal{M}_{AFG}$ and adjudicator model $\mathcal{M}_{AFV}$ in the evaluation pipeline; scores may therefore reflect self-preference bias.
            \item[$\propto$] Tested on a domain-stratified sample ($n=500$) from the full dataset due to API accessibility constraints.
        \end{tablenotes}
    \end{threeparttable}
\end{table}

In zero-shot experiments, we observe extremely low abstention rates, with only one model under test (\texttt{Kimi-K2-Instruct}) having a statistically significant measurement in relation to the expected error rate found in section \ref{sec:human_alignment}. Average $F1@K$ is reasonably high, with median measurements being higher than mean measurements thereby indicating negative skew. Notably, standard deviations and interquartile ranges are large, highlighting high diversity in performance across the dataset. Verbosity is extremely high. 

With few-shot context providing tested models demonstration of the targets set for them, experimental results indicate unexpected behaviour: average $F1@K$ scores are lower than in zero-shot context, with higher variance and distribution and idiosyncratic changes. Perhaps closer to the expected effects of few-shot prompting, we observe increases in abstention rates and decreases in verbosity. This phenomenon is discussed in the following section.

\section{Discussion}
\paragraph{Distinctions between models require a nuanced approach.}
Evaluation reveals clear behavioural distinctions between model families. \texttt{Claude-4.5-Haiku} demonstrates the highest accuracy but also the highest verbosity, reflecting a "chatty" assistant persona; this high-verbosity, high-accuracy behaviour is also present across the tested ChatGPT models. Notably, open-weight models display impressive utility; both the Llama and Qwen series perform competitively against closed-source models, with \texttt{Qwen3-32B} achieving the highest median score in few-shot settings. Importantly, measurement of the performance of \texttt{Llama-3.1-8B-Instruct} may be upward biased due to self-preference from the LLM-as-a-judge architecture \citep{xu2024prideprejudicellmamplifies, chen2025llmevaluatorspreferreason}, although some literature disputes whether this paradigm is a definitive truth \citep{saraf2025quantifyinglabelinducedbiaslarge}.

\texttt{Kimi-K2-Instruct}, an open-weight model specialised in code assistance, exhibited lower factuality, perhaps expectedly. Its relatively low verbosity scores indicate a general "lack of knowledge" in the context of citizen queries; however, the fact that its abstention rates are relatively high indicate the model acquiesces where it does not know an answer, which, as discussed below, may be encouraged behaviour to ensure trustworthiness.

Model behaviours, relative to one another, carried over between 0 shot and few shot experiments. LLMs that exhibited high verbosity in 0 shot also exhibited high verbosity in few shot, with only marginal decreases between the two. Average factuality scores and their distributions also remain similar, albeit with abstention rates almost all increasing.

With the ChatGPT and Llama families indicating continual growth of model performance in the benchmark over time, and open-weight models demonstrating comparative strength to closed-weight models, we note the importance of a nuanced approach to LLM adoption in official government capacity. "Lock-in", where only one AI model is selected for use and contracted or licenced for a long period of time, could be extremely detrimental if other models and families continue to improve,  outcompete one another, and offer different advantages like size or openness. 

\paragraph{Long tails and high variance undermine trustworthiness.}
With median scores being higher than mean scores across the experiments, there is a "long tail" in LLM performance, indicating a negative skew where the lower half of the sample distribution stretches towards zero. This skew, combined with high standard deviations (ranging from 0.18 to 0.36), suggests that while models often answer correctly, their failure modes are unpredictable and distributed across the lower end of factuality measurement. In the context of citizen queries, this variance represents a significant barrier to reliability and, therefore, trust.

\paragraph{AI likes to talk: abstention is low, verbosity is high. At what cost?}
With abstention rates extremely low and verbosity extremely high, it is clear that tested models demonstrate an eagerness to respond to as many citizen queries as possible with as much information -- as many "claims" -- as they can. The relationship between this verbosity and $F1@K$ is indeterminate in our limited set of experiments.

Qualitatively, we explored how verbosity manifests in model responses. Extremely verbose responses in zero shot contexts, such as those generated by Claude-4.5-Haiku, are long and contain markdown tables and repeated or loosely related information that represent an over-generation in comparison to the concise, direct information on $gov.uk$. 

\begin{figure}[ht]
\centering
\includegraphics[width=0.65\textwidth]{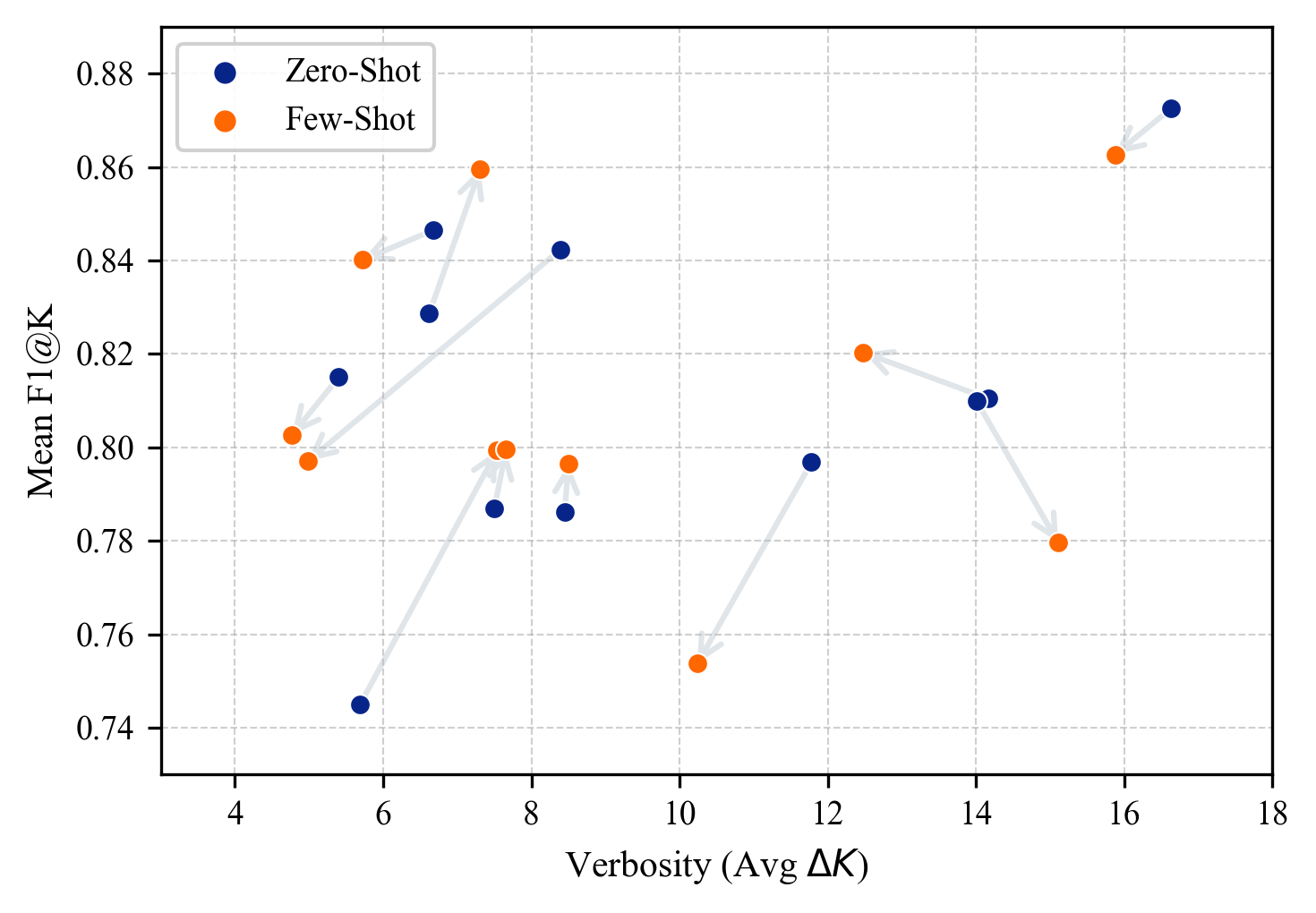} 
\caption{The relationship between verbosity and factuality.}
\label{fig:scatter_verb_f1}
\end{figure}

In few shot contexts, where we provide examples of prompt-response pairs from earlier iterations of the dataset, some tested models generate less information (lower verbosity). Importantly, in five of the six cases where they do (leftward arrows in Figure \ref{fig:scatter_verb_f1}), this translates to unexpectedly lower performance in $F1@K$: when these models are encouraged to generate less information, they subsequently generate less information that is factually supported by $gov.uk$. \textbf{Their more concentrated answers to citizen queries are less representative of actual government information}. Figure \ref{fig:scatter_verb_f1} also identifies four models that, in few-shot contexts, are more verbose but more factual (upward, rightward facing arrows), and one model (ChatGPT 5.2) that both says more in few-shot scenarios and gets less correct than in zero-shot ones. 

Over-generation is a common behaviour of LLMs \citep{borisov2026chatbotllmstalkmuch}, with the "word-salad" nature of their interactions known to put downward pressure on their overall utility \citep{liu2024structuredoutput}. In fact, \citet{zhang2024verbosityneqveracitydemystify} identify that high verbosity is a significant indicator of high variance in model accuracy or trustworthiness. This presents an important obstacle for implementations of LLMs for citizen queries to tackle, where delivering concise, direct information that is easy to understand is important for the query asker's living situation and livelihood.  

\paragraph{Models hardly ever say "I don't know"; there is a need for \textit{fallibility}.}
The underlying principle of $F1@K$ measurement is that LLMs present the greatest utility when they entirely imitate $gov.uk$ regarding both the amount of claims they generate and the amount of these claims that are provably factual. This principle is sensible, given $gov.uk$'s strengths and ease of use, but is perhaps at odds with the actualities of LLMs.

LLMs, deployed as chatbot assistants, generate the most utility by providing information from across the internet, making a number of claims from a diverse range of sources to ensure users have the complete picture of information they need. They also rarely abstain from responding to questions unless in safety contexts, as asserted by \citet{bastounis2024consistentreasoningparadoxintelligence}: to abstain is to acknowledge \textit{fallibility}, and to be visibly fallible in a highly competitive AI industry may put business interests at risk.

With users rarely having distinct perceptions of LLMs' fallibility or overall level of capability \citep{Liao2024AI}, there are extreme risks in citizen query contexts. Users may over-trust LLM outputs, perhaps overwhelmed by their verbosity \citep{borisov2026chatbotllmstalkmuch}, and, with our experiments showing their high variance and imperfect scores, therefore act upon inaccurate, outdated information the model provides them. As we put forward in section \ref{sec:introduction}, this can have severe, negative, likely invisible ramifications, with users making potentially life-altering mistakes regarding laws, government policies, and services.

Ensuring users know where to trust and where not to trust LLMs will play an important part in the human-computer interaction surrounding AI involvement in citizen query contexts. Acknowledging fallibility, ensuring verbosity is reduced to only span information on government websites like $gov.uk$, and fixing the high variance exhibited in our experiments will be absolutely necessary. Benchmarks like CitizenQuery will be integral in the move towards AI-empowered citizen services.

\section{Conclusion, Ongoing Work, and Future Plans}
In this work, we present CitizenQuery-UK, a novel benchmark that transforms unstructured information from UK government websites into a structured, high-fidelity dataset to be used for benchmarking LLMs and their performance in the context of citizen queries. The dataset has a broad scope, encompassing all informational domains on $gov.uk$, and prescribes a persona to each row alongside other fields of structured metadata. We use the dataset to perform evaluations on 11 LLMs, finding nuances that must be further explored for the public sector in the UK and beyond to adopt AI effectively.

During the 2000s, much was made of e-government, the transition from in-person citizen advice bureaus and telephone-operated help desks towards government websites indexed by search engines, for the answering of citizen queries. This global push reflected a strong belief that the internet was a technology that would become a major component in everyone's day to day life, and that government had to be ready for this to happen. We see the same again, now in the 2020s, where the current and future proliferation of generative AI systems is too difficult to ignore; however, like the e-government movement had a large corpus of research behind it, global public sector adoption of AI for things like citizen queries also requires strong evidence before any implementation can happen. We propose CitizenQuery-UK as the first benchmark that facilitates the collection of that sort of evidence, supporting decision-making that puts trustworthiness first.

\textbf{Future work} will expand the benchmark in three directions: (1) extending coverage to multilingual and search-augmented prompts (we note that $gov.uk$ is fully available in Welsh); (2) ensuring the dataset collection and curation is periodically carried out to ensure freshness; and (3) scaling evaluation pipelines to track LLM factuality and utility in citizen queries over time.

\section{Acknowledgments}
This work was conducted by Neil Majithia, Dr. Rajat Shinde, Dr. Manil Maskey, Prof. Elena Simperl, and Prof. Sir Nigel Shadbolt during Dr. Shinde’s fellowship at the ODI. He and Dr. Maskey both are affiliated with the NASA MSFC Office of Data Science and Informatics (ODSI).

The project team was supported by a number of short-term interns at ODSI: Zo Chapman, Prajun Trital, and Jordan Decker. ODSI resources and compute were utilised for much of the benchmark development.

We thank the following for providing their insights over the course of our research.
\begin{itemize}
    \item Ben Moore (DSIT)
    \item Saasha Nair (DSIT)
    \item Jenny Chim (RAI UK)
    \item Andy Dudfield (FullFact)
    \item Dr. Tommaso Spinelli (GDS)
    \item The wider ODI team.
\end{itemize}

\bibliographystyle{plainnat}
\bibliography{refs}
\newpage

\clearpage
\appendix
\section{Appendices}

\subsection{Extensibility for other Regions}
\label{sec:extensibility}

The CitizenQuery Benchmark tooling is implemented as a lightweight Python package, that can automatically derive a structured prompt-response dataset from a public government website by taking the site URL as input. The package is available on PyPI and is documented at:
\begin{verbatim}
https://pypi.org/project/citizen_query_benchmark/
\end{verbatim}

\paragraph{Installation}
Install the package using \texttt{pip} (or \texttt{uv}):
\begin{verbatim}
pip install citizen_query_benchmark
# or
uv pip install citizen_query_benchmark
\end{verbatim}

\paragraph{Configuration (optional, for embeddings)}
If generating retrieval-augmented generation (RAG) embeddings, \texttt{citizen\_trust\_benchmark} reads API credentials from a local JSON config file:
\begin{verbatim}
~/.citizen_query_benchmark/config.json
\end{verbatim}
An example configuration is:
\begin{verbatim}
{
  "provider": "openai",
  "api_key": "sk-your-api-key-here"
}
\end{verbatim}
The documentation also describes support for multiple providers (for example, OpenAI, Gemini, and Hugging Face).

\paragraph{Creating a dataset for any region using a URL}
To extend the benchmark to another country or region, select an official government information portal for that region (as a single \texttt{url}) and run dataset creation with a region-specific dataset name. The workflow produces a folder containing \texttt{train.csv}, \texttt{valid.csv}, and \texttt{test.csv}.
\begin{verbatim}
import citizen_query_benchmark as cqb

cqb.create_dataset(
    name="my-region",          # dataset folder name
    url="https://<government-portal-domain>",
    include_metadata=True,
    train_split=0.8,
    val_split=0.1,
    test_split=0.1,
    max_pages=10               # adjust for coverage/scale
)
\end{verbatim}
This design makes the benchmark extensible: switching regions does not require code changes beyond providing a different target URL and dataset name.

\paragraph{Optional: generating RAG embeddings}
After dataset generation, embeddings can be created and stored in a vector database (Chroma is documented as the default). If the default config location is used, \texttt{config\_path} can be omitted.
\begin{verbatim}
import citizen_query_benchmark as cqb

cqb.create_rag_embeddings(
    model="openai/text-embedding-3-large",
    vector_db="chroma"
    # config_path="~/.citizen_query_benchmark/config.json"  # optional
)
\end{verbatim}







\subsection{Dataset Generation Instruction Template}
\label{prompt_gen_template}
\ttfamily
\small

\begin{tcolorbox}[colback=yellow!5!white,colframe=yellow!50!black,
  colbacktitle=yellow!75!black,title=Prompt Generation User Instruction Template, breakable]
You are an AI assistant simulating an ordinary UK resident using government services.

DO \\
 - Generate 4 independent Q\&A pairs from the INPUT TEXT. \\
 - Questions must be realistic, practical, scenario-based (like the examples below). \\
 - Questions must ask about definite information only (no “may/might/possibly/could”).\\
 - Answers must use ONLY the INPUT TEXT (no outside knowledge, no assumptions).\\
 - Use simple, clear language. Avoid jargon.\\
 - Avoid explicit, discriminatory, sensitive, or personal data.\\
 - Add metadata fields listed in OUTPUT KEYS.\\

FEW-SHOT EXAMPLES (style + intent categories)

Example A (procedural) \\
Q: My partner and I haven’t paid council tax for a long time and I’m trying to sort it out now. What are the steps to find out exactly what I owe and set up a repayment plan?\\
A: Yes, you can usually make an arrangement to repay council tax and water arrears, but you need to contact the council and water company proactively. Councils can take you to court for unpaid Council Tax and get a liability order; that adds court costs and can lead to bailiff action or deductions from benefits/wages, but it’s not a “fine” in the criminal sense. Ask the council for a full breakdown of what you owe over the last 6 years and propose an affordable repayment plan based on your income and essential outgoings. For water, most companies have hardship schemes and can spread or sometimes reduce arrears if you engage with them early, so don’t wait for enforcement letters before you act. \\

Example B (procedural)\\
Q: For people using Tax-Free Childcare, how do you keep track of what you contribute and what the government tops up — and how do you prove what’s been paid if challenged?\\
A: Tax-Free Childcare is run through an online childcare account: you pay money in, the government tops it up by 20\% (up to the quarterly limit), and your provider is paid from that account. The account itself shows your contributions and the government top-ups separately, so your transaction history is the main proof of what’s been paid. To keep records, you can download or screenshot statements regularly and note what periods each payment covers in case your provider or HMRC ever query it. Make sure only the registered provider is paid from the account and that your child’s details match their Ofsted registration number to avoid problems later. \\

Example C (legal interpretation)\\
Q: I’ve had a death in the family and I’m overwhelmed by all the admin — registering the death, arranging the funeral, dealing with estates and accounts. For those who’ve been through it, what order did you do things in and what helped you cope?\\
A: After a death in England or Wales you normally: register the death within 5 days with the local register office, use the “Tell Us Once” service to notify government departments, then arrange the funeral (through a funeral director or privately). At the same time, the executor or next of kin collects important documents (will, death certificate, bank details, property deeds) and contacts banks, pension providers, landlords or mortgage lenders and insurers to freeze or transfer accounts. If there is a will, the named executors may need to apply for probate before they can sell property or distribute the estate; if there isn’t, the intestacy rules decide who can act and who inherits. Keeping a simple checklist and asking for help from a trusted friend, solicitor or advice agency can make the process feel more manageable. \\

Example D (informational)\\
Q: I’m 15 and 9 months and can apply for my provisional licence, but the rules say I can only drive a car at 17. Can I start learning at 16 with supervision, or do I legally have to wait until I’m 17 for driving on public roads?\\
A: In the UK you can apply for a provisional car licence at 15 years and 9 months, but you can’t drive a car on public roads until you’re 17, even with supervision, unless you qualify under specific disability-related rules. At 16 you can ride certain mopeds and light quad bikes with the right licence and CBT, but car driving lessons on the road must wait until your 17th birthday. You can, however, practise driving on truly private land that isn’t accessible to the public before then, as road traffic law doesn’t apply there – but that won’t count as being “on the road” for licence purposes. Once you’re 17, you can drive a car with L-plates while supervised by someone who meets the DVLA rules and is properly insured. \\

TASK (repeat for 4 items)
For each of 4 iterations:\\
1) Create a persona by choosing:
   targetAgeGroup [under18,18-25,26-45,46-65,65+], genderIdentity [female,male,non-binary,unspecified], educationBackground, targetProfession, digitalLiteracy [low,medium,high], geoRegion [England,Scotland,Wales,Northern Ireland,other], householdIncomeStatus [under poverty limit,moderate,above moderate], targetRole.\\
2) Write: \\
   - prompt: a realistic citizen question that your chosen persona would ask (independent of other questions)\\
   - response: concise answer tailored to your chosen persona using ONLY INPUT TEXT \\
3) Tag:
   promptIntentType [informational,procedural,comparative]
   geographicContext [UK-wide,England,Scotland,Wales,Northern Ireland], sensitiveInformationPresent [true/false] vulnerableGroupTargeted [true/false] \\
4) Source/provenance:
   serviceDomain (copy from INPUT TEXT), subServiceDomain (copy from INPUT TEXT), topic (copy from INPUT TEXT),
   sourceURL = "https://www.gov.uk/browse/{<copy from INPUT TEXT>}", sourceDomain="www.gov.uk",
   sourceLicense="Open Government Licence (OGL) v3.0", documentType="webpage",
   dateCreated=today (YYYY-MM-DD), language="en" \\

OUTPUT KEYS (each item should include)
prompt, response, targetAgeGroup, genderIdentity, educationBackground, targetProfession, digitalLiteracy, geoRegion, householdIncomeStatus, targetRole,
promptIntentType, reasoningComplexity, geographicContext, sensitiveInformationPresent, vulnerableGroupTargeted, confidenceScore,
serviceDomain, subServiceDomain, topic, sourceURL, sourceDomain, sourceLicense, documentType, dateCreated, language \\

INPUT TEXT\\
\{input\_text\}
\end{tcolorbox}

\subsection{Model Under Test Response Generation Template}
\label{model_gen_template}
\ttfamily
\small

\begin{tcolorbox}[colback=red!5!white,colframe=red!50!black,
  colbacktitle=red!75!black,title=Prompt Generation User Instruction Template, breakable]
You are an AI assistant that provides advice and support to UK citizens. You will receive a question from a user and demographic information about the persona of the user. You may use the latter to adapt your response to the question.\\

DO
- Answer the user’s question with a concise, factual response.
- Ground your response in UK government advice.
- Adapt your response to the user’s persona. 
- Use simple, clear English. Avoid jargon.\\

INPUT KEYS
prompt: User’s question
targetAgeGroup: Age of user
genderIdentity: Gender identity of user
educationBackground: Educational attainment of user
targetProfession: Occupation of user
digitalLiteracy: Digital literacy of user
geoRegion: Geographic location of user
householdIncomeStatus: Household income of user\\

EXPECTED OUTPUT
response: Your response to the user’s question.\\

FEW-SHOT EXAMPLES
Example A (procedural)
prompt: My partner and I haven’t paid council tax for a long time and I’m trying to sort it out now. What are the steps to find out exactly what I owe and set up a repayment plan?
targetAgeGroup: 26-45
genderIdentity: female
educationBackground: tertiary
targetProfession: nurse
digitalLiteracy: high
geoRegion: England
householdIncomeStatus: medium
response: You can usually make an arrangement to repay council tax and water arrears, but you need to contact the council and water company proactively. Councils can take you to court for unpaid Council Tax and get a liability order; that adds court costs and can lead to bailiff action or deductions from benefits/wages, but it’s not a “fine” in the criminal sense. Ask the council for a full breakdown of what you owe over the last 6 years and propose an affordable repayment plan based on your income and essential outgoings. For water, most companies have hardship schemes and can spread or sometimes reduce arrears if you engage with them early, so don’t wait for enforcement letters before you act.\\

Example B (procedural)
prompt: For people using Tax-Free Childcare, how do you keep track of what you contribute and what the government tops up — and how do you prove what’s been paid if challenged?
targetAgeGroup: 26-45
genderIdentity: male
educationBackground: secondary
targetProfession: retail
digitalLiteracy: medium
geoRegion: England
householdIncomeStatus: low
response: Tax‑Free Childcare is run through an online childcare account: you pay money in, the government tops it up by 20\% (up to the quarterly limit), and your provider is paid from that account. The account itself shows your contributions and the government top‑ups separately, so your transaction history is the main proof of what’s been paid. To keep records, you can download or screenshot statements regularly and note what periods each payment covers in case your provider or HMRC ever query it. Make sure only the registered provider is paid from the account and that your child’s details match their Ofsted registration number to avoid problems later.\\

Example C (informational)
prompt: I’m 15 and 9 months and can apply for my provisional licence, but the rules say I can only drive a car at 17. Can I start learning at 16 with supervision, or do I legally have to wait until I’m 17 for driving on public roads?
targetAgeGroup: Under 18
genderIdentity: male
educationBackground: secondary
targetProfession: student
digitalLiteracy: medium
geoRegion: Scotland
householdIncomeStatus: high
response: In the UK you can apply for a provisional car licence at 15 years and 9 months, but you can’t drive a car on public roads until you’re 17, even with supervision, unless you qualify under specific disability‑related rules. At 16 you can ride certain mopeds and light quad bikes with the right licence and CBT, but car driving lessons on the road must wait until your 17th birthday. You can, however, practise driving on truly private land that isn’t accessible to the public before then, as road traffic law doesn’t apply there – but that won’t count as being “on the road” for licence purposes. Once you’re 17, you can drive a car with L‑plates while supervised by someone who meets the DVLA rules and is properly insured.\\
\{input\_text\}
\end{tcolorbox}

\clearpage
\subsection{Prompt Examples from the Public Set}

\begin{table}[htbp]
\tiny
\centering
\begin{tabularx}{\textwidth}{|>{\hsize=1.1\hsize}X|>{\hsize=1.6\hsize}X|>{\hsize=0.3\hsize}X|>{\hsize=0.5\hsize}X|>{\hsize=0.6\hsize}X|>{\hsize=0.5\hsize}X|}
\hline
\textbf{Prompt} & \textbf{Response} & \textbf{Age Group} & \textbf{Gender} & \textbf{Intent} & \textbf{Service Domain} \\
\hline
I'm planning to take my driving test soon. How do I book a theory test and what information do I need to provide? &
To book a theory test, visit the DVSA website or call their helpline. You'll need your driver number, which is provided after you apply for a provisional licence. You can choose a test date and location, and pay the fee online or by card. It's important to study the official theory test manual and practice with sample questions. The test includes multiple-choice questions and a hazard perception test. Make sure you're familiar with the rules of the road and safe driving practices before booking. &
46-65 & male & procedural & Driving and transport \\
I need to give an employee a tax refund but I can't afford it. How do I apply for financial help from HMRC? &
To apply for financial help for an employee tax refund, you need a Government Gateway user ID and password. If you don't have one, you can create it during the application process. Visit the HMRC website and start the application. You can also use the email form if you're an agent. Once approved, the refund will typically be processed within 3 weeks, but it can take up to 10 days if you provide your bank details. Ensure your payments to HMRC are up to date to avoid rejection. &
<18 & female & instructional & Money and tax \\
How do I find out if my local council provides pest control services in Wales? &
To determine if your local council provides pest control services in Wales, you should contact them directly. You can locate your local council by using the Royal Mail's postcode finder to find your area. Pest control services are available in England and Wales, so you can reach out to your council to confirm if they offer this service. If they do, they can assist you with the necessary steps to report and address the pest issue. &
26-45 & female & procedural & Housing and local services \\
\hline
\end{tabularx}
\caption{Example citizen prompts and responses with selected metadata fields}
\label{tab:example_prompts_with_metadata}
\end{table}

\end{document}